\documentclass[doublecol]{epl2}
\usepackage{amsmath}
\usepackage{amsfonts}
\usepackage{amssymb}
\usepackage{color}
\usepackage{ulem}
\usepackage{bm}
\usepackage{graphicx}

\newcommand{\mathsym}[1]{{}}
\newcommand{\unicode}[1]{{}}

\newcommand{\hide}[1]{}
\newcommand{\blue}[1]{\textcolor{blue}{#1}}

\newcommand{\figref}[1]{Fig.~\ref{#1}}

\newcommand{\equref}[1]{Eq.~(\ref{#1})}
\newcommand{\tauLJ}{\tau_\mathrm{LJ}}
\newcommand{\ta}{t_\mathrm{age}}

\newcommand{\tquench}{\Delta t_{\mathrm{quench}}}

\newcommand{\delete}[1]{}

\title{Aging in amorphous solids: A study of the first passage time and persistence time distributions}

\author{Nima H. Siboni$^{1}$ \and Dierk Raabe$^{1}$ \and Fathollah Varnik$^{2}$\thanks{corresponding author: \email{fathollah.varnik@rub.de}}}
\institute{
	\inst{1} Max-Planck Institut f\"ur Eisenforschung, Max-Planck Stra{\ss}e~1, 40237 D\"usseldorf, Germany\\
	\inst{2} Interdisciplinary Centre for Advanced Materials Simulation (ICAMS), Ruhr-Universit\"at Bochum, Universit\"atsstra{\ss}e 150, 44780 Bochum, Germany}

\pacs{83.80.Hj}{Suspensions, dispersions, pastes, slurries, colloids}
\pacs{05.40.-a}{Fluctuation phenomena, random processes, noise, and Brownian motion}

\abstract{
The time distribution of relaxation events in an aging system is investigated via molecular dynamics simulations. The focus is on the distribution functions of the first passage time, $p_1(\Delta t)$, and the persistence time, $p(\tau)$.  In contrast to previous reports, both $p_1$ and $p$ are found to evolve with time upon aging. The age dependence of the persistence time distribution is shown to be sensitive to the details of the algorithm used to extract it from particle trajectories. By updating the reference point in event detection algorithm and accounting for the event specific aging time, we uncover age the dependence of $p(\tau)$, hidden to previous studies. Moreover, the apparent age-dependence of $p_1$ in continuous time random walk with an age independent $p(\tau)$ is shown to result from an implicit synchronization of all the random walkers at the starting time.}
\begin{document}

\maketitle

\section{Introduction}

The nature of the glassy state has been the subject of intense studies over the past decades (see, e.g., \cite{Barrat2002,Cipelletti2005,Berthier2011} and references therein). Due to time evolution towards an ideally unreachable equilibrium state, properties of a glass do not obey time translation invariance but depend on the aging time $\ta$, the time elapsed between the quench (from the liquid to the glassy phase) and the beginning of the measurement. This \delete{age dependence} is best seen in dynamical properties \cite{Kob1997a} but also shows itself in the mechanical response of the system such as the peak stress during shear start-up \cite{Varnik2004}.

An interesting observable for the study of the age-dependence is the probability distribution function of the \textit{persistence time}, $\tau$, which measures the time between two successive relaxation events. The aging behavior of the persistence time distribution function, $p(\tau)$, provides useful information about the underlying physical mechanism. For example, if $p(\tau)$ becomes broader without changing its shape, this may hint towards the fact that aging slows down all the microscopic processes with the same rate. A change in the shape of $p(\tau)$, on the other hand, would indicate a modification of relaxation channels upon aging.

A quantity, closely related to $p(\tau)$, is the distribution function for the so-called first passage time, $p_1(\Delta t)$, where $\Delta t$ is the time interval between the beginning of the measurement and the first relaxation event. The terminology used here is based on Refs.~\cite{Warren2009,Warren2010a,Warren2010b,Warren2013}. Other authors have used other expressions. For example, the persistence time, $\tau$, is called 'waiting time' in~\cite{Heuer2008} and 'exchange time' in~\cite{Jung2005,Hedges2007}. In the latter references, the first passage time, $\Delta t$, is referred to as 'persistence time'. For a $p(\tau)$ with a finite first moment, $p_1(\Delta t)$ is uniquely determined via \cite{Feller1957,Barkai2003},
\begin{equation}
p_1(\Delta t)=\frac{\int_{\Delta t}^{\infty} p(\tau)d\tau}{\overline{\tau}}.
\label{eq:p1-equil}
\end{equation}
An important consequence of \equref{eq:p1-equil} is that, an age-independent persistence time distribution will always lead to an age-independent distribution of the first passage time,  as long as $\overline{\tau}=\int p(\tau)d\tau$ (a measure of the relaxation time) is finite. Surprisingly, in recent molecular dynamics (MD) simulations, aging effects are observed in $p_1$, while $p(\tau)$ seems to be age-independent~\cite{Warren2009,Warren2010a,Warren2010b,Warren2013,Vollmayr-Lee2013}. This behavior has been attributed to a violation of \equref{eq:p1-equil} in the case of a diverging $\overline\tau$ \cite{Warren2009,Warren2010a,Warren2010b,Warren2013,Vollmayr-Lee2013}. However, it is generally known that the average relaxation time $\overline{\tau}$ in a system quenched into a glassy state grows continuously  with the system age, $\ta$ , remaining finite for a finite $\ta$~\cite{Warren2013}. It is, therefore, worth asking why aging effects are not observed in the persistence time distribution function in MD simulations of aging systems. The present letter addresses this issue.

For the purpose of this study, exactly the same simulation model as in~\cite{Warren2009} is chosen. In agreement with \cite{Warren2009,Warren2010a,Warren2010b}, a marked dependence of the first passage time distribution function on $\ta$ is observed. In contrast to \cite{Warren2009,Warren2010a,Warren2010b}, however, our simulations clearly reveal the age dependence of $p(\tau)$ as expected from \equref{eq:p1-equil}. The sampling process for the persistence time, $\tau$, starts only after the detection of the first relaxation event. This introduces an additional aging time of $\Delta t$, during which the system dynamics slows down further. Resolving aging effects on $p(\tau)$ thus requires a survey of the system dynamics over longer times than in the case of $p_1(\Delta t)$. Consequently, with the same numerical effort, a less pronounced aging effect is observed in $p(\tau)$ as compared to $p_1(\Delta t)$. Moreover, $p(\tau)$ is quite sensitive to the details of the event detection algorithm. We propose two slight, yet important, improvements here. Firstly, in previous works, the reference point in determining the displacement vector of a particle is fixed once for all times. The detection of a given jump then depends on the relative orientation of the new and old displacement vectors so that a significant number of jumps may remain undetected (\figref{typDis1}). This problem can be solved by shifting the reference point to the position of the last jump. Secondly, since $\tau$ is the time difference between two successive events, $\tau=t_{i+1}-t_i$, we propose to account for the additional aging until the time $t_i$. 

We also address the question why in standard continuous time random walk models (CTRW)~\cite{Bouchaud1992,Monthus1996,Barkai2003} an age-dependent $p_1$ may occur even if $p(\tau)$ is age-independent. This is shown to originate from an implicit synchronization of trajectories, as $t=0$ corresponds to the occurrence of an event. Such a feature is absent in MD simulations, where the origin of time marks a random point between two relaxation events.

\section{Detection of the relaxation events}\label{sec:algorithm}
In amorphous solids, a structural relaxation event can be defined in various ways. Common approaches make use of (i) the collective motion of particles in the configuration phase space \cite{Stillinger1995, Stillinger1982, Doliwa2003}, (ii) the motion of a particle relative to its neighbors  \cite{Falk1998,Goldhirsch2002,Rabani1997}, or (iii) the single particle displacement \cite{Lindemann1910}. It has recently been shown that these criteria provide similar results for the distribution of relaxation events \cite{Ahn2013}.

\begin{figure} 
\centering
\hide{
\begin{tikzpicture}[scale=2]
\draw[->,>=latex,thick,blue] (0,0) -- (1,0);
\draw[->,>=latex,thick,blue] (1,0) -- (1.17,0.98);
\draw[->,>=latex,thick,blue] (1.17,0.98) -- (1.17+0.98,0.98+0.17);
\draw[->,>=latex,thick,black] (0.,0) -- (2.15,1.15);
\draw[-,>=latex,dotted,black] (2.15,1.15) -- (2.15+0.88,1.15+0.469);
\draw[->,>=latex,dashed,blue] (2.15,1.15) -- (2.15+0.98,1.15-0.17);
\draw[->,>=latex,dashed,black] (0,0) -- (3.13,0.98);
\node[thick] at (2.1,0.4) {$\Delta\bm{r}(t_{n+1})$};
\node[thick] at (.75,0.6) {$\Delta\bm{r}(t_{n})$};
\node[thick] at (2.41,1.2) {$\theta$};
\node[thick] at (2.15+0.4,1.15-0.14) {$\blue{\delta \bm{r}}$};
\end{tikzpicture}
}
\includegraphics[width=0.35\textwidth]{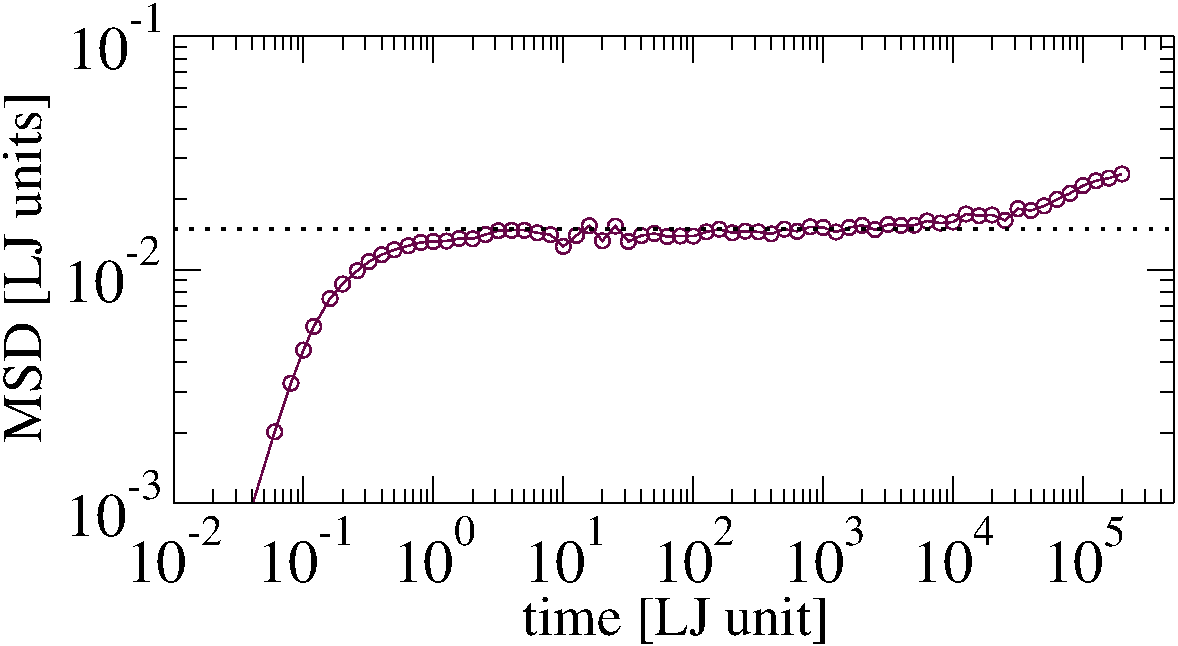}\\
(a)~~~~~~~~~~~~~~~~~~~~~~~~~~~~~~~~~~~~~~~~~~~~~~~~~~~~~~~~~~~~~~~~\\
~~~~~~~~\includegraphics[width=0.3\textwidth]{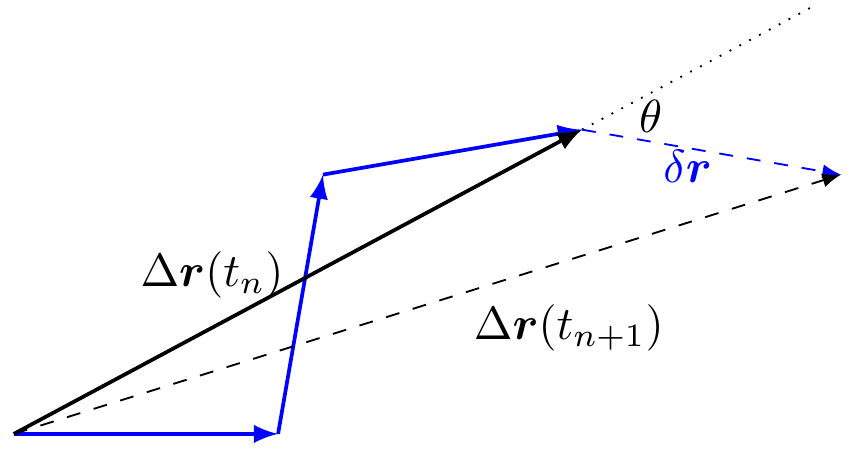}\\
(b)~~~~~~~~~~~~~~~~~~~~~~~~~~~~~~~~~~~~~~~~~~~~~~~~~~~~~~~~~~~~~~~~\\
(c)\label{schematic_subfig_c}
\includegraphics[width=0.4\textwidth]{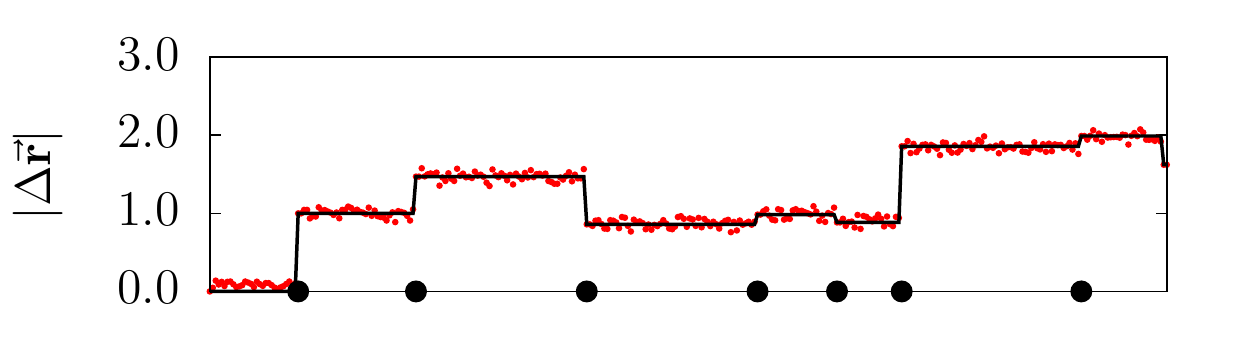}\\
(d)
\includegraphics[width=0.4\textwidth]{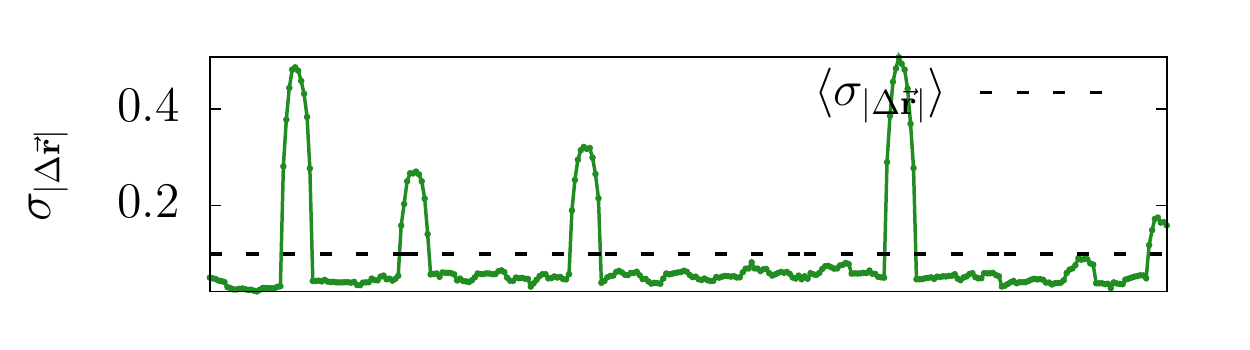}\\
(e)
\includegraphics[width=0.4\textwidth]{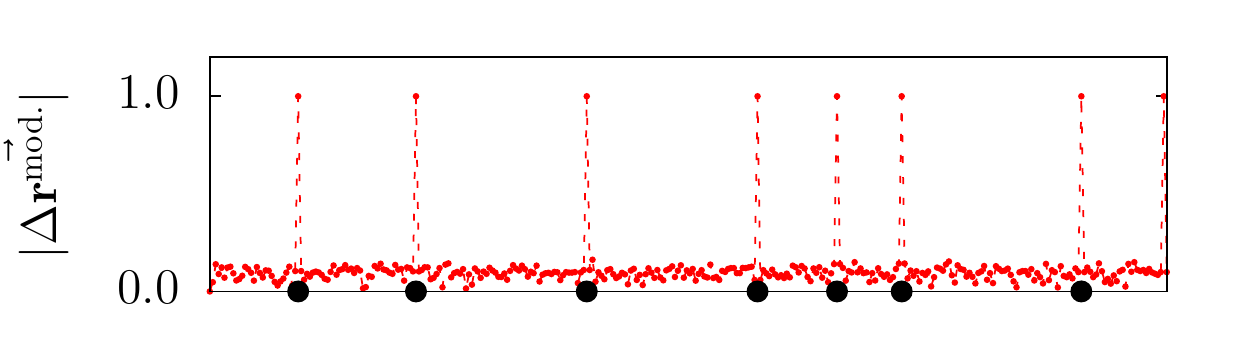}\\
(f)
\includegraphics[width=0.4\textwidth]{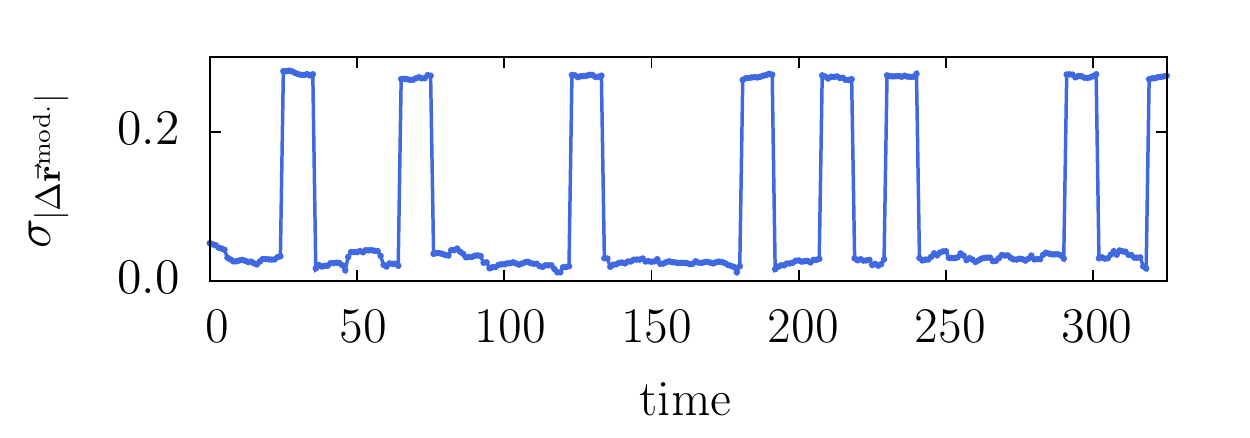}\\
\caption{(a) Mean square displacement for a binary LJ glass for the aging time of $\ta=2\times 10^5$ ($T=0.2$). The dotted line indicates the Lindemann criterion for cage breakage \cite{Lindemann1910}. A particle is said to escape from the cage made by its neighbors as its mean square displacement exceeds this (empirical) threshold. (b) A schematic particle trajectory made of successive jumps, each jump corresponding to a cage relaxation event. In (c), a typical one dimensional particle displacement is constructed as a combination of small scale rattling motion with larger scale jumps. The solid line shows the displacement without the small scale fluctuations. The black circles mark the cage-breakage events. Although the size of all the larger scale jumps is the same, the standard deviation of displacements misses some of the cage relaxation events if the displacement vector is calculated with respect to a fixed point, $\Delta\bm{r}=\bm{r}(t)-\bm{r}(0)$. Shifting the reference point after each jump to the new particle position, i.e., using $\Delta\bm{r}^{\mathrm{mod.}}=\bm{r}(t)-\bm{r}(t_{n-1})$ allows to detect all the relaxation events (e,f).}
\label{typDis1}
\end{figure}
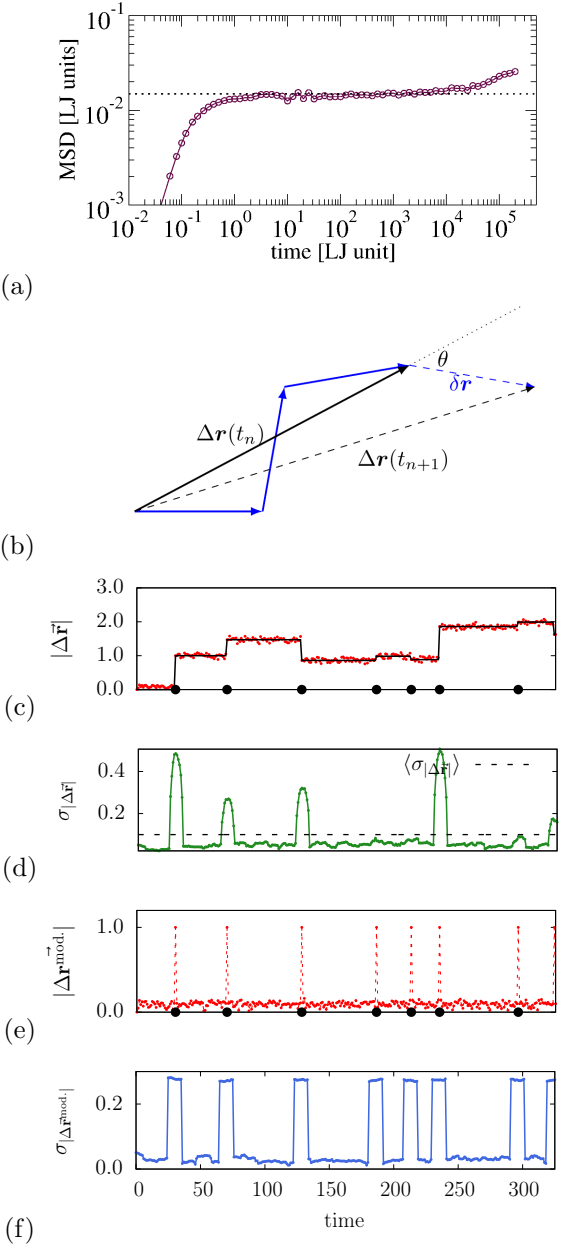

Typical dynamics of a particle in a glassy state consists of in-cage high-frequency rattling motion accompanied by intermittent cage-breakage jumps (structural relaxation events). A way to detect relaxation events is to monitor the magnitude of individual particle's displacement, $|\Delta {\bm{r}}|=|\bm{r}(t)-\bm{r}(0)|$, as a function of time and to determine the corresponding standard deviation $\sigma_{|\Delta \bm{r}|}$ over a sliding time window  
\begin{align}
\sigma_{|\Delta \bm{r}|}=\langle|\Delta {\bm{r}}|^2\rangle-\langle|\Delta {\bm{r}}|\rangle^2,
\end{align} where $\langle ... \rangle$ represents time average over the sliding time window (the use of the absolute magnitude ensures that $\langle|\Delta {\bm{r}}|\rangle \ne 0$). The point at which  $\sigma_{|\Delta \bm{r}|}$ acquires a peak indicates a jump. For a time window within which the particle only rattles in its cage, the standard deviation reflects the amplitude of the in-cage vibrations. As the sliding time window covers a jump, the standard deviation increases sharply, reflecting particle motion of the order of the cage size. This increase is eventually followed by a decrease, as the sliding time window passes over the event, and only covers the rattling motion of the particle in its new cage. Each peak of $\sigma_{|\Delta \bm{r}|}$ thus marks a jump. To illustrate this, we generate a random trajectory, $\bm{r}(t)=\bm{r}(0)+\sum_k \delta \bm{r}_k$, with small $|\delta \bm{r}_k|$, characteristic of rattling motion, followed by, less frequent, larger displacements to mimic cage breakage. An example for $|\Delta \bm{r}(t)| = |\sum_{t_k\le t} \delta \bm{r}_k|$ obtained from such a trajectory is shown in \figref{typDis1} (c). The peaks in the resulting standard deviation, $\sigma_{|\Delta \bm{r}|}$, reflect jumps in $|\Delta \bm{r}|$ (\figref{typDis1} (d)). Quantitatively, we identify jumps via $\sigma_{|\Delta \bm{r}|}>\left<\sigma_{|\Delta \bm{r}|}\right>$, where $\left< \cdots \right>$ stands for the average over the entire simulation time. Introducing such a threshold is necessary due to the stochastic nature of the rattling motion which leads to fluctuations of $\sigma_{|\Delta \bm{r}|}$.

Although all jumps detected by this method correspond to jumps in the particle's trajectory, a number of jumps remain undetected. This is due to relatively small changes of the $|\Delta \bm{r}|$ for some of the jumps. As illustrated in \figref{typDis1} (c), the change in $|\Delta\bm{r}|$ for the $4^{\mathrm{th}}$, $5^{\mathrm{th}}$ and $7^{\mathrm{th}}$ jumps is relatively small, and the corresponding $\sigma_{|\Delta \bm{r}|}$-peaks remain undetected. Note that, in this example, the magnitude of all the large scale $\delta \bm{r}_k$ (associated with a cage breakage event) is deliberately set to a constant value (of $1$). Thus, the failure of the method is not related to the size of the step. To see the main reason, we recall the Cauchy inequality, $|\Delta\bm{r}+\delta\bm{r}|-|\Delta\bm{r}|\le |\delta\bm{r}|$. The equality applies if $\delta\bm{r}$ is parallel to $\Delta\bm{r}$ and in the trivial case of $\Delta\bm{r}=\bm{0}$ (\figref{typDis1} (b)).

As done in MD studies of supercooled liquids~\cite{Hedges2007}, a simple remedy to this problem is to ensure $\Delta\bm{r}_{\mathrm{previous~displacements}}=\bm{0}$ via shifting the reference point from $\bm{r}(0)$ to the last detected jump, 
\begin{align}
\Delta\bm{r}^{\mathrm{mod.}}(t)=\bm{r}(t)-\bm{r}(t_{n}),\;\;\;\;\;\;\; t_n < t \le t_{n+1}.
\end{align}

Using this modified definition, and the corresponding modification to the standard deviation,
\begin{align}
\sigma_{|\Delta \bm{r}^{\mathrm{mod.}}|}=\langle|\Delta {\bm{r}^{\mathrm{mod.}}}|^2\rangle-\langle|\Delta {\bm{r}^{\mathrm{mod.}}}|\rangle^2,
\end{align} we analyze the same random trajectory as used in \figref{typDis1} (c-d). As shown in \figref{typDis1} (e), all the jumps are clearly visible in the behavior of $\Delta\bm{r}$  and find their corresponding peaks in the standard deviation (\figref{typDis1}(f)). Even though the trajectory chosen here is schematic, the related analysis provides important insight into the proper jump detection algorithm. In the present MD simulations, the number of the detected cage breakage events reduces by roughly a factor of two, if the aforementioned modification is not used. Note, however, that this issue is irrelevant for the first passage time, since $\bm{r}(0)$ is the only meaningful reference point here.

\section{Results}\label{sec:results}

\subsection{Simulation setup}\label{TheBinaryLennardJonesModel}
We use the well-known 80:20 binary Lennard-Jones (LJ) mixture \cite{Kob1995,Kob1995a}, which has proven to be a suitable model for the study of various aspects of the glass transition and the response of glassy systems to an external perturbation (see, e.g., \cite{Varnik2003,Baschnagel2005,Berthier2002a,Warren2009} and references therein). The LJ particles (types  A and B) interact via
$U_{\text{LJ}}(r)= 4\epsilon_{\alpha\beta}[(d_{\alpha\beta}/r)^{12}-(d_{\alpha\beta}/r)^6]$ 
with $\alpha,\beta={\text{A,B}}$, $\epsilon_{\text{AB}}=1.5\epsilon_{\text{AA}}$, $\epsilon_{\text{BB}}= 0.5\epsilon_{\text{AA}}$, $d_{\text{AB}}= 0.8d_{\text{AA}}$, $d_{\text{BB}}= 0.88d_{\text{AA}}$. The masses of type A and B are equal, $m_{\text{A}}= m_{\text{B}}$. The potential is truncated at twice the minimum position of the LJ potential, $r_{\text{c},\alpha\beta} = 2.245 d_{\alpha\beta}$. The parameters $\epsilon_{\text{AA}}$, $d_{\text{AA}}$ and $m_{\text{A}}$ define the units of energy, length and mass. The unit of time is given by $\tauLJ = d_{\text{AA}}\sqrt{m_{\text{A}} / \epsilon_{\text{AA}}}$. The simulation box is a three dimensional cube of length $L=10$ with periodic boundary conditions along $x,y$ and $z$ directions. The particle number density is $\rho=1.2$. All the simulations reported here are performed at constant volume ($NVT$-ensemble). Equations of motion are integrated using the velocity-Verlet combined with the Nos\'e-Hoover algorithm, with time discretization $\delta t=0.005$.

The system is first equilibrated at $T=4$ (liquid state) and then cooled to $T=0.3$ (glass) at a constant rate. Once this final temperature is reached, $T$ is kept fixed for the rest of the simulation. Two quench durations of $\tquench=100$ (fast quench) and $750$ (slow quench) are used, corresponding to cooling rates of $\dot\Gamma=3.7\times 10^{-2}$ and $\dot\Gamma=5\times 10^{-3}$, respectively. The sample preparation protocol with the slow quench is identical to the conditions implemented in Ref.~\cite{Warren2009}, which enables us to compare our results with that reference. The study of the fast quench rate is motivated by the more prominent aging effects, which simplify its detection. To have reliable statistics, $40$ independent simulations are performed for each set of parameters.

\subsection{First passage time and persistence time}\label{FirstPassageTimeAndPersistenceTime}
In all the simulations reported below, the end of the quenched process is considered as the origin of the time ($t=0$). The system is then let to evolve with time at constant temperature for a duration of $\ta$. The detection of the structural events starts at time $t=\ta$ by following the trajectories of the individual particles. If a particle undergoes its first relaxation event at time $t_{1}$, the corresponding first passage time is obtained as $\Delta t = t_1 -\ta $.

The next relaxation event for the same particle provides the persistence time $\tau=t_{2}-t_{1}$. In contrast to the first passage time, where all the events correspond to the same aging time, it is $t=t_1=\ta+\Delta t$ which marks the beginning of the detection process to capture the second event. More generally, to a persistence time between the events $t_{n}$ and $t_{n+1}$, there corresponds the age $t_{n}$. It is noted that even the smallest correction of $\Delta t$ might change the age-dependence of $p(\tau)$ significantly as the first passage time can be comparable to $\ta$ \cite{Fielding2000,Barkai2003}. We remark that our definition is in line with the age interpretation in~\cite{Vollmayr-Lee2013}.

\begin{figure}
\centering
(a)\includegraphics[width=0.4\textwidth]{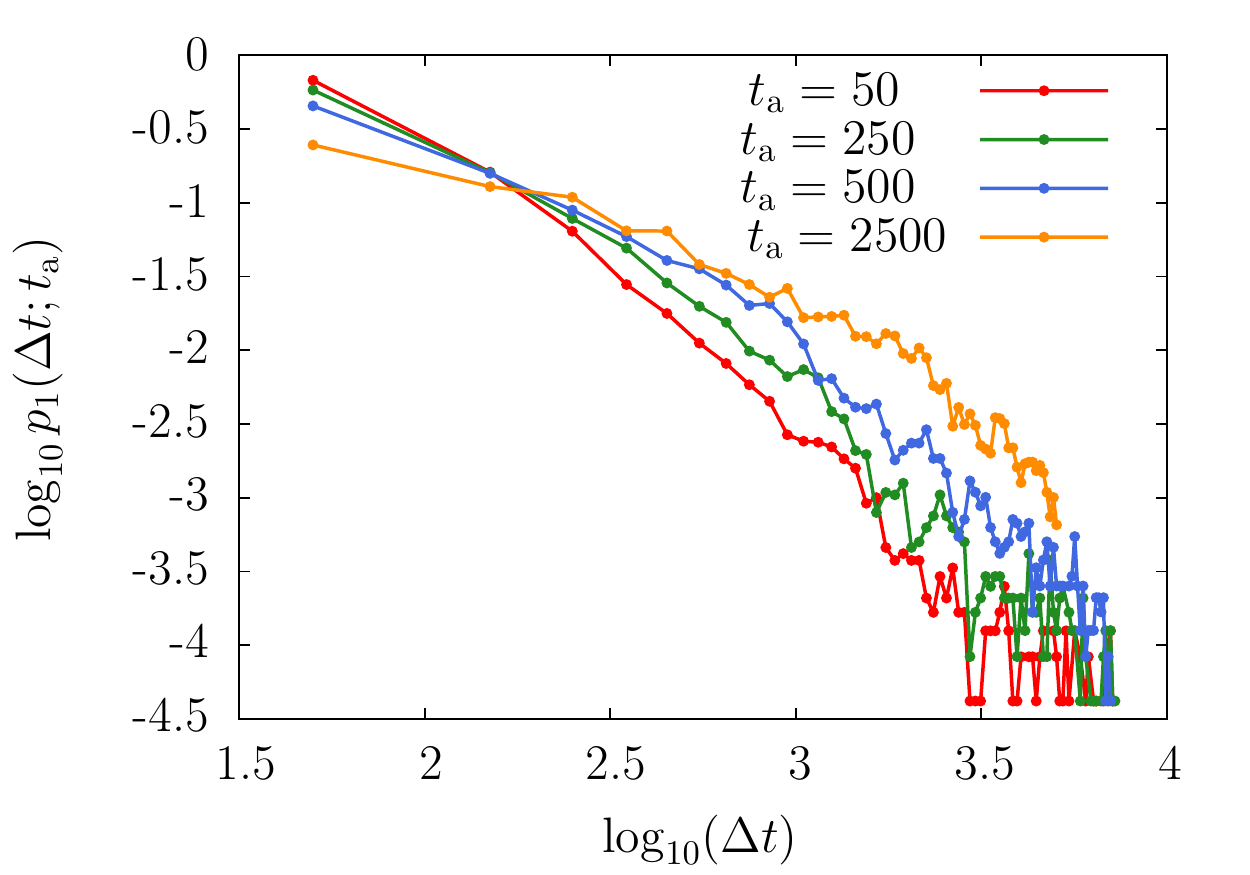}
(b)\includegraphics[width=0.4\textwidth]{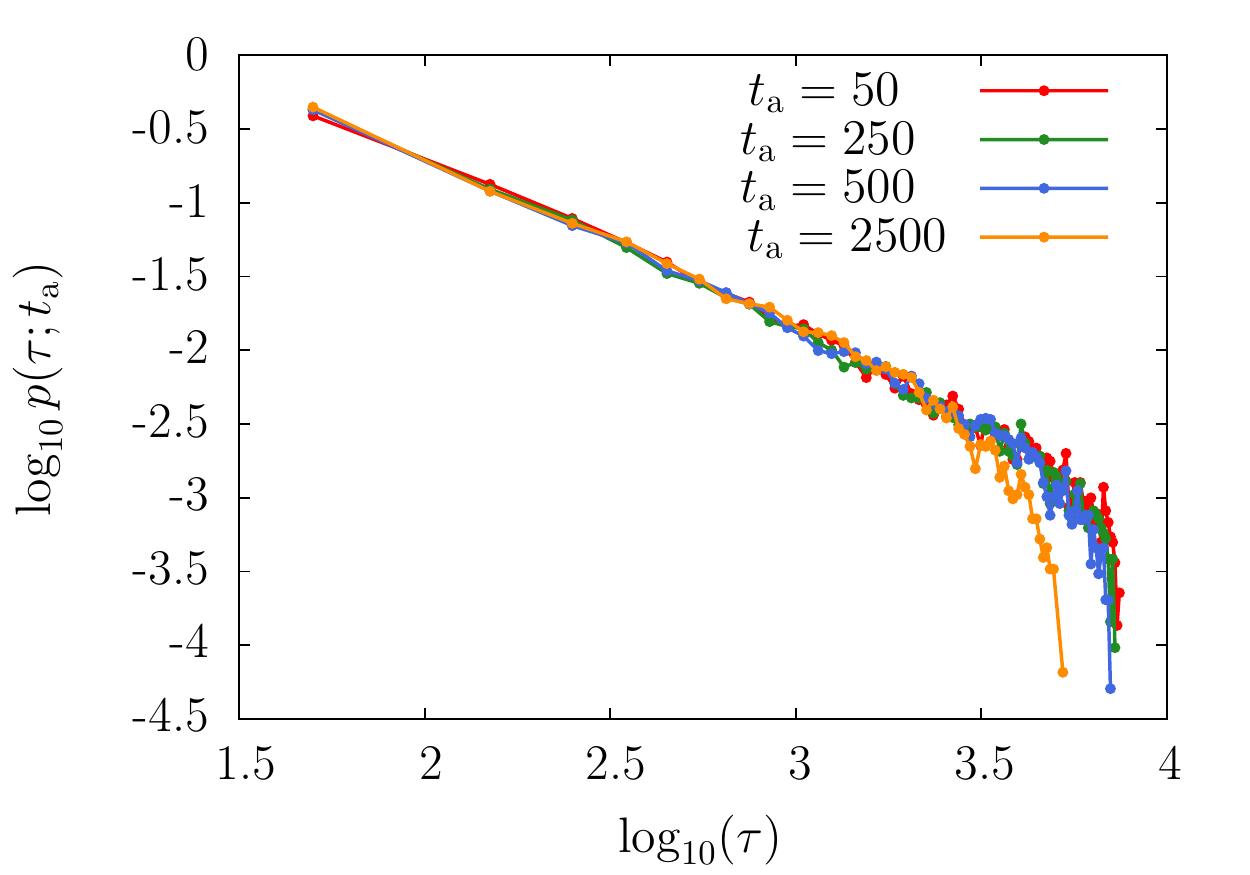}
(c)\includegraphics[width=0.4\textwidth]{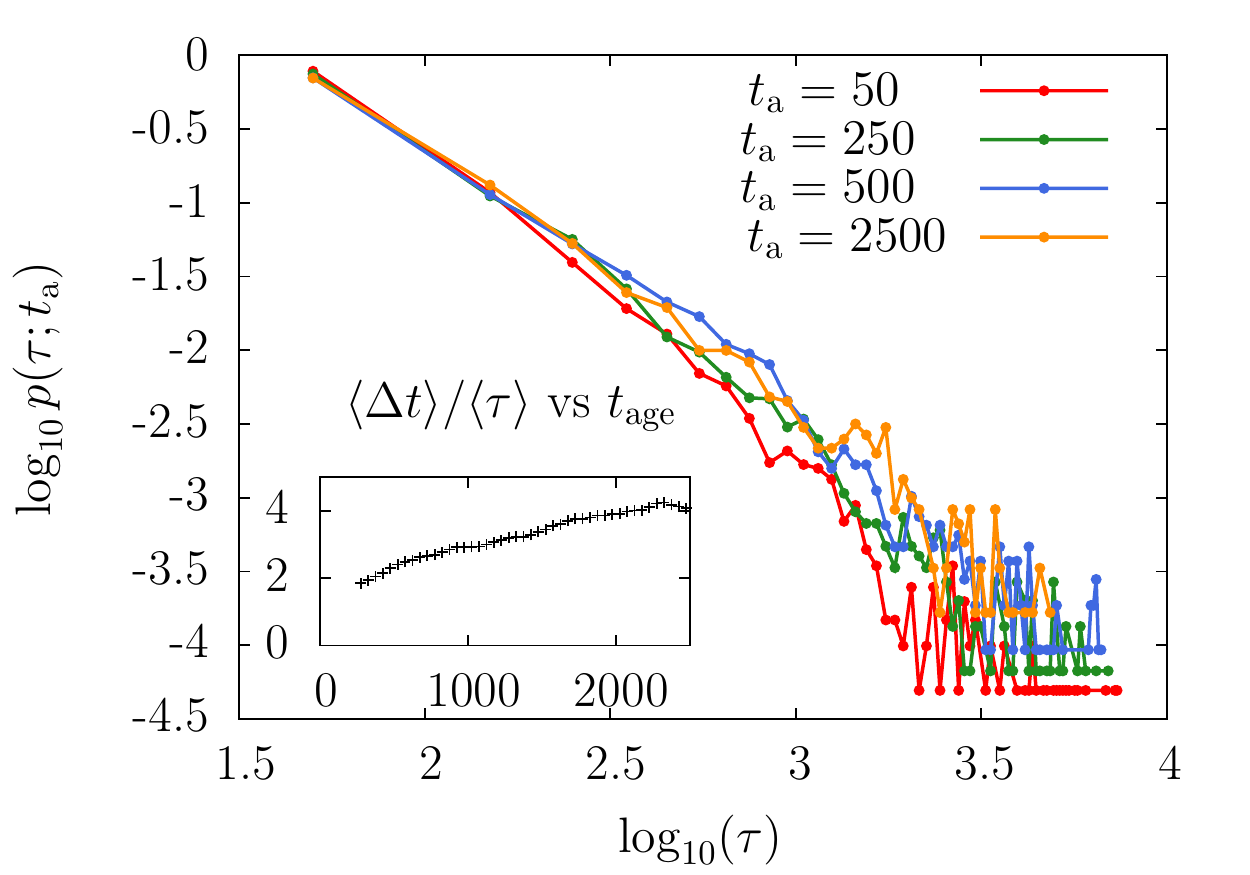}
\caption{(a) The first passage and (b) the persistence time distribution functions for a binary LJ mixture, determined after cooling from $T=4$ to $T=0.3$ with a cooling rate of $\dot\Gamma=5 \times 10^{-3}$ and a subsequent aging process. Different curves correspond to different aging times $\ta$ as indicated. 
The details of the simulation and the jump detection algorithm are identical to~\cite{Warren2009}. The panel (c) shows $p(\tau)$ determined using the modified algorithm proposed in this work. The aging effect hidden in panel (b) is now revealed.	The inset of panel (c) depicts the ratio of the average first passage time, $\langle\Delta t\rangle$, to the average persistence time, $\langle\tau\rangle$, as a function of $\ta$.}
\label{fig:p1+ptau-slow}
\end{figure}

\begin{figure}
\centering
(a)\includegraphics[width=0.4\textwidth]{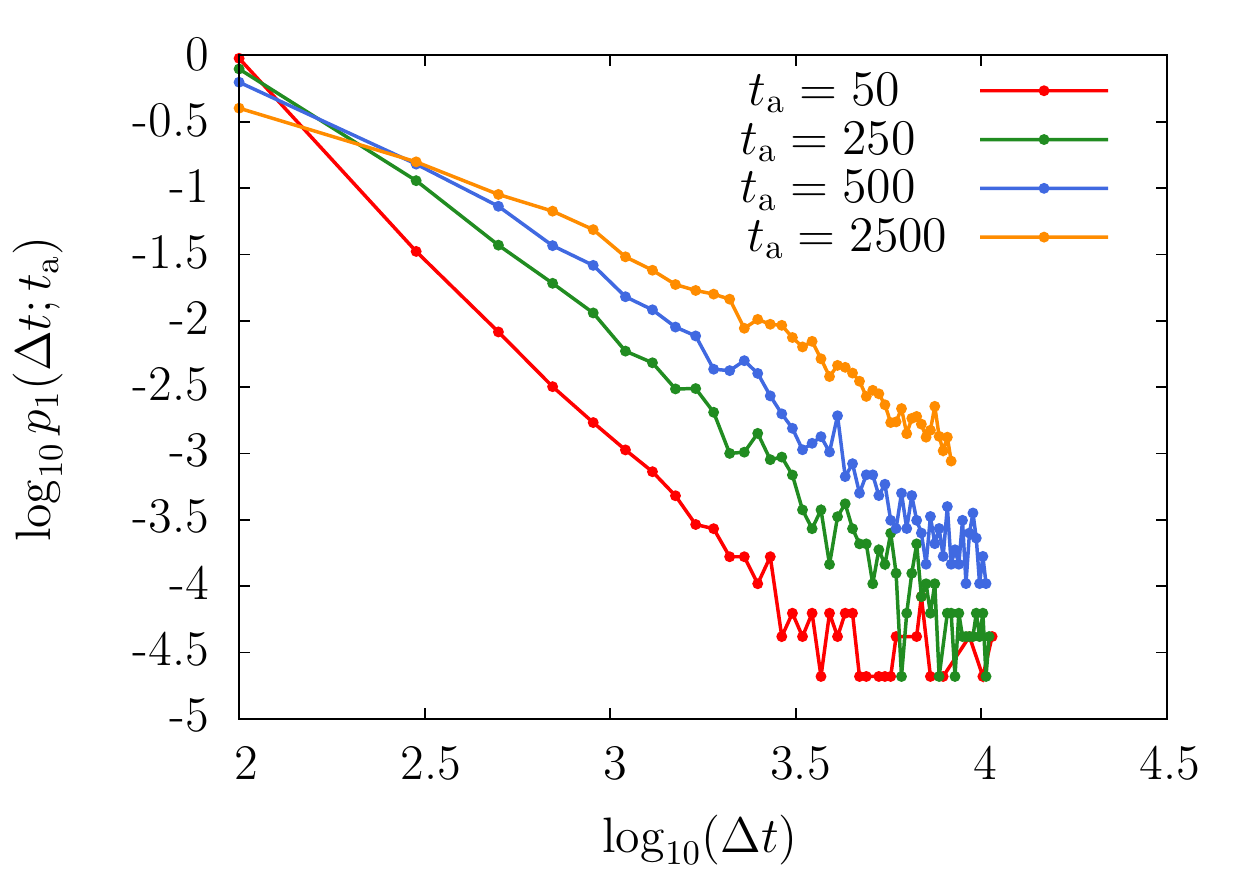}
(b)\includegraphics[width=0.4\textwidth]{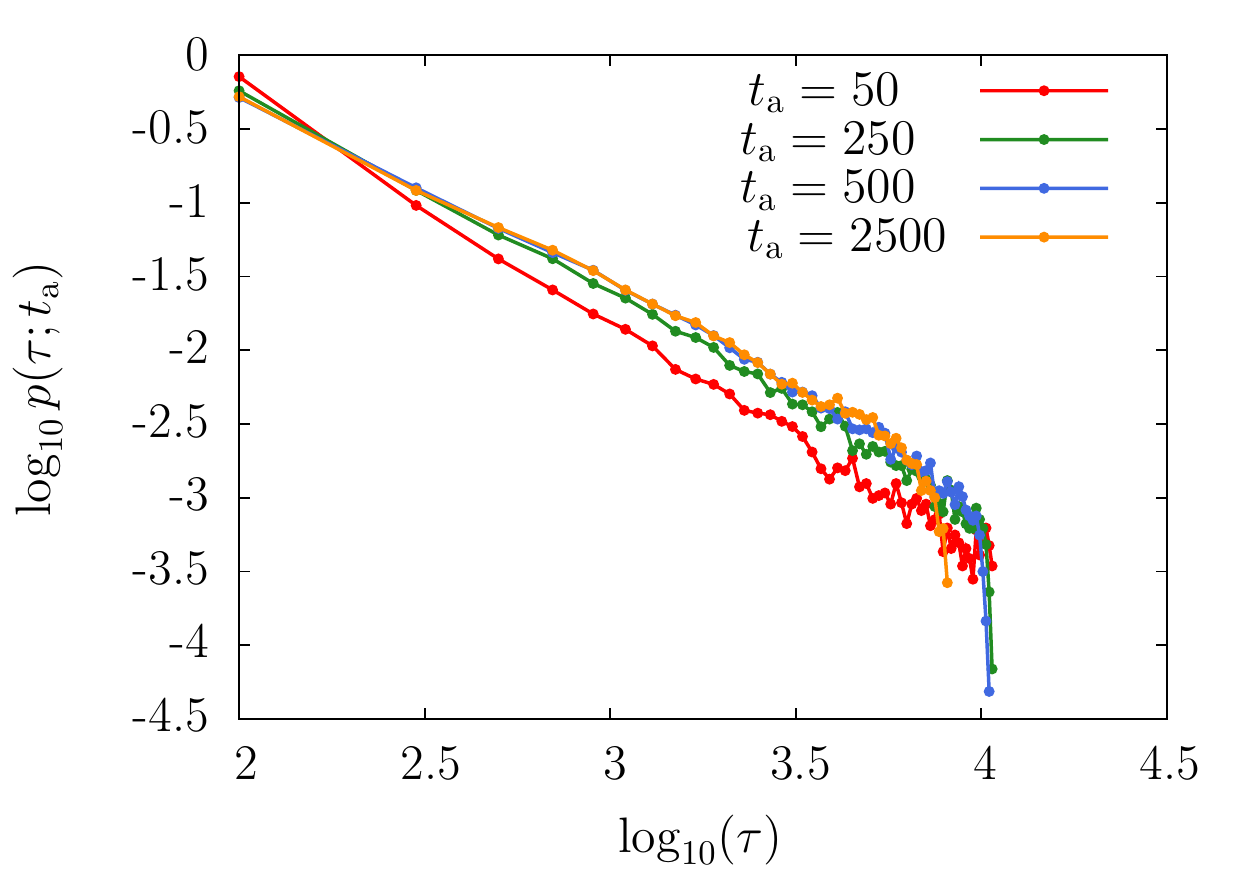}
(c)\includegraphics[width=0.4\textwidth]{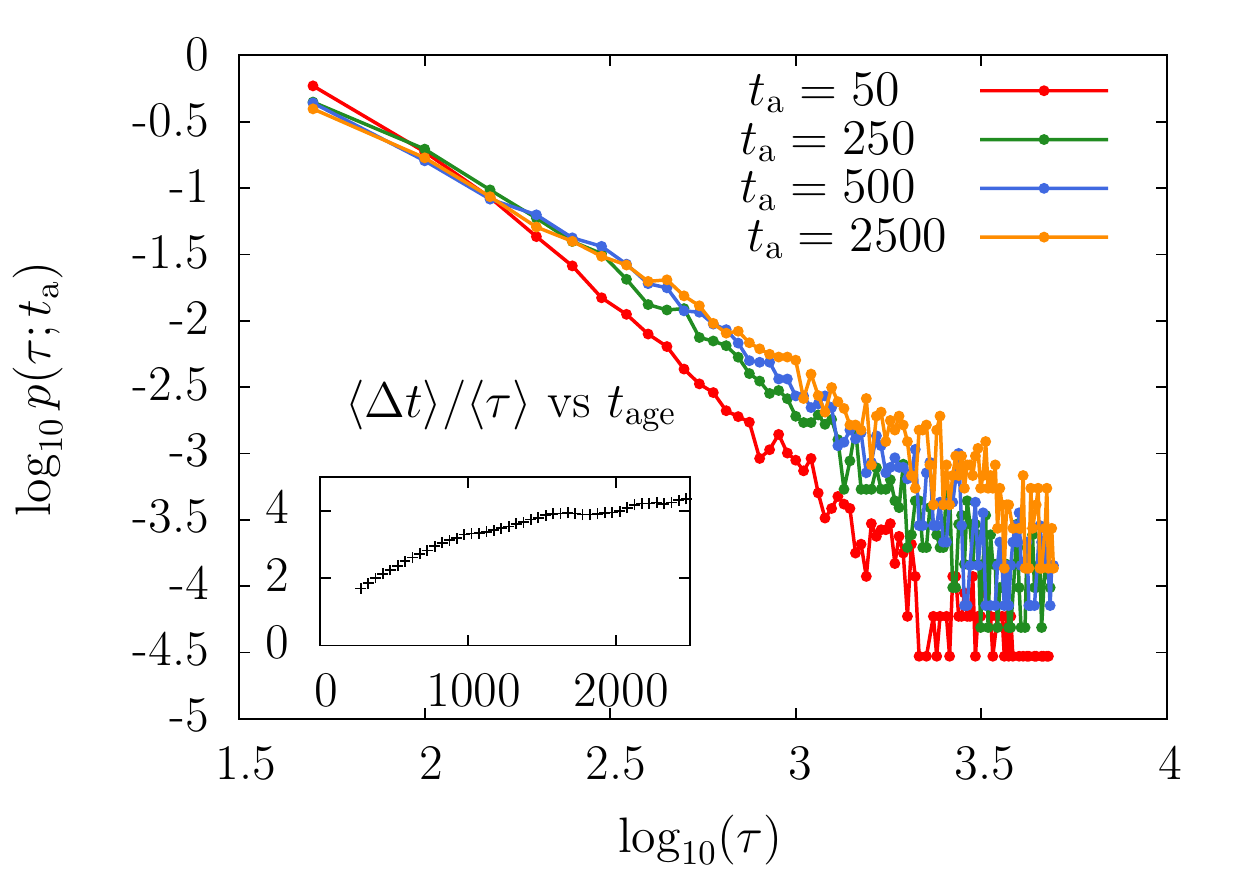}
\caption{The same data as in \figref{fig:p1+ptau-slow} but for a higher cooling rate of $\dot\Gamma=3.7 \times 10^{-2}$. Now aging effects in the persistence time distribution are strong enough to become visible regardless of the details of the detection algorithm.}
\label{fig:p1+ptau-fast}
\end{figure}

Simulation results on the first passage time and the persistence time distribution functions are shown in \figref{fig:p1+ptau-slow}. Results shown in the panels (a) and (b) of this figure are in perfect agreement with that reported in~\cite{Warren2009}. In particular, the first passage time shows a marked dependence upon aging (panel a) while the persistence time distribution remains hardly affected by $\ta$ (panel b). In contrast to \figref{fig:p1+ptau-slow}b, a clear age-dependence of $p(\tau)$ is visible in \figref{fig:p1+ptau-slow}c. The data shown in this panel are obtained for exactly the same trajectories as used in the panel (b) but with the new algorithm proposed in this work. Interestingly, the ratio of the average first passage and persistence times shows an age dependence, reminiscent of a similar trend in supercooled liquids upon temperature variation (inset in Figs.~\ref{fig:p1+ptau-slow}c and ~\ref{fig:p1+ptau-fast}c)~\cite{Hedges2007}.

In order to highlight the age-dependence of the persistence time distribution function further, we have also performed simulations at a higher cooling rate of $\dot\Gamma=3.7 \times 10^{-2}$. As seen from \figref{fig:p1+ptau-fast}b, aging effects are now sufficiently pronounced to be observable even without the corrections proposed in the present paper. The age-dependence of $p(\tau)$ is, however, better resolved in \figref{fig:p1+ptau-fast}c, where the same trajectories are analyzed using the new approach.

\section{Aging in the continuous time random walk}\label{TrapModelCTRWandACTRW}

As a generalization of the simple random walk, the step size and the waiting time
(time interval between two successive steps) are not constant in a CTRW but are drawn from the corresponding distribution functions\footnote{The step size distribution function appears not to be of significant importance in the case of the glassy materials \cite{Doliwa2003}. Therefore, here we will focus on the time probability distribution function.}. 

In the context of modeling dynamic processes in glass forming systems \cite{Monthus1996,Doliwa2003,Rubner2008,Helfferich2014a}, the waiting time $\tau$ for an activated process is often written as $\tau=\Gamma_0^{-1} e^{{E}/{k_B T}}$. Here, $E$ is the energy barrier, $\Gamma_0$ is the attempt rate, and $T$ is the (effective) temperature~\cite{Ilg2007,Nicolas2014}. In amorphous systems, the energy barrier does not have a single value but is characterized by a probability distribution function, $\rho(E)$, which leads to a distribution function for the waiting or persistence time, $p(\tau)$.

Interestingly, the model shows aging in the first passage time distribution $p_1(\Delta t)$, in spite of an age independent $p(\tau)$. As will be shown shortly, this is a consequence of an implicit synchronization of CTRW-trajectories. In MD simulations, the beginning of the sampling process does not necessarily coincide with the occurrence of a relaxation event and, therefore, this type of aging behavior is not expected to occur.

\begin{figure}
\hide{
\begin{tikzpicture}[scale=3]
\draw[|-,blue] (-5,0)--(-4,0);
\draw[|-,red] (-4,0)--(-2.5,0);
\draw[|-] (-2.5,0)--(-1.0,0);
\draw[|-, dotted] (-1.0,0)--(0.7,0);
\draw[|-,blue] (0.7,0)--(2.7,0);
\draw[|-] (2.7,0)--(3.7,0);
\draw[dotted, |-] (3.7,0)--(4.5,0);
\draw[thick] (-5,-0.5) --  (-5,+0.5);
\node[thick] at (-5,-0.75) {$t=0$};
\draw[thick] (-2.2,-0.5) --  (-2.2,+0.5);
\node[thick] at (-2.2,-0.75) {$t_{\mathrm{a,1}}$};
\node[thick] at (-1.5,-0.4) {$\Delta t$};
\draw[-|,dashed] (-2.2,-0.2)--(-0.99,-0.2);
\draw[thick] (2.0,-0.5) --  (2.0,+0.5);
\node[thick] at (2.0,-0.75) {$t_{\mathrm{a,2}}\gg t_{\mathrm{a,1}}$};
\node[thick] at (2.4,-0.4) {$\Delta t$};
\draw[-|, dashed] (2.0,-0.2)--(2.71,-0.2);
\node[thick,blue] at (-4.5,0.2) {$\tau_1$};
\node[thick,red] at (-3.25,0.2) {$\tau_2$};
\node[thick,green] at (1.7,0.2) {$\tau_i$};
\node[thick] at (3.25,0.2) {$\tau_{i+1}$};
\end{tikzpicture}
}
\includegraphics[width=0.4\textwidth]{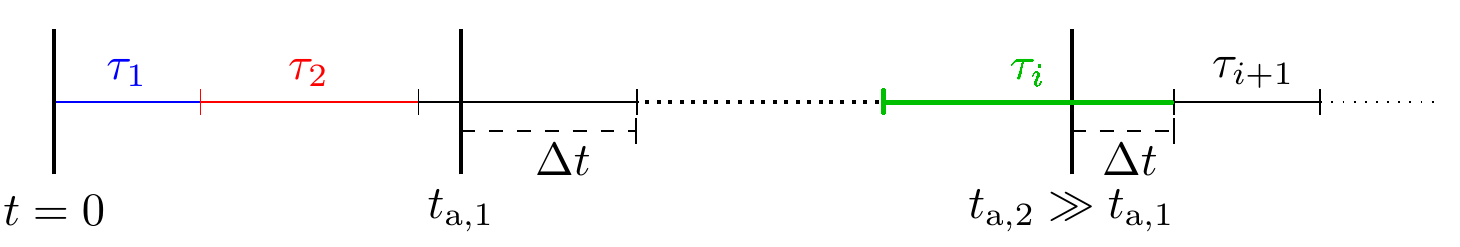}
\caption[]{Illustration of the CTRW along the time axis. Starting at $t=0$, one draws successively waiting (or persistence) times, $\tau_1$,  $\tau_2,\, \ldots,$ from a given persistence time distribution $p(\tau)$. The first passage time, $\Delta t$, is then determined by taking the first event after an aging time of $\ta$. The probability distribution of $\Delta t$ is obtained by repeating this procedure.}
\label{fig:CTRW-schematic}
\end{figure}
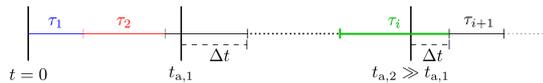

\begin{figure}
\centering
(a)\includegraphics[width=0.4\textwidth]{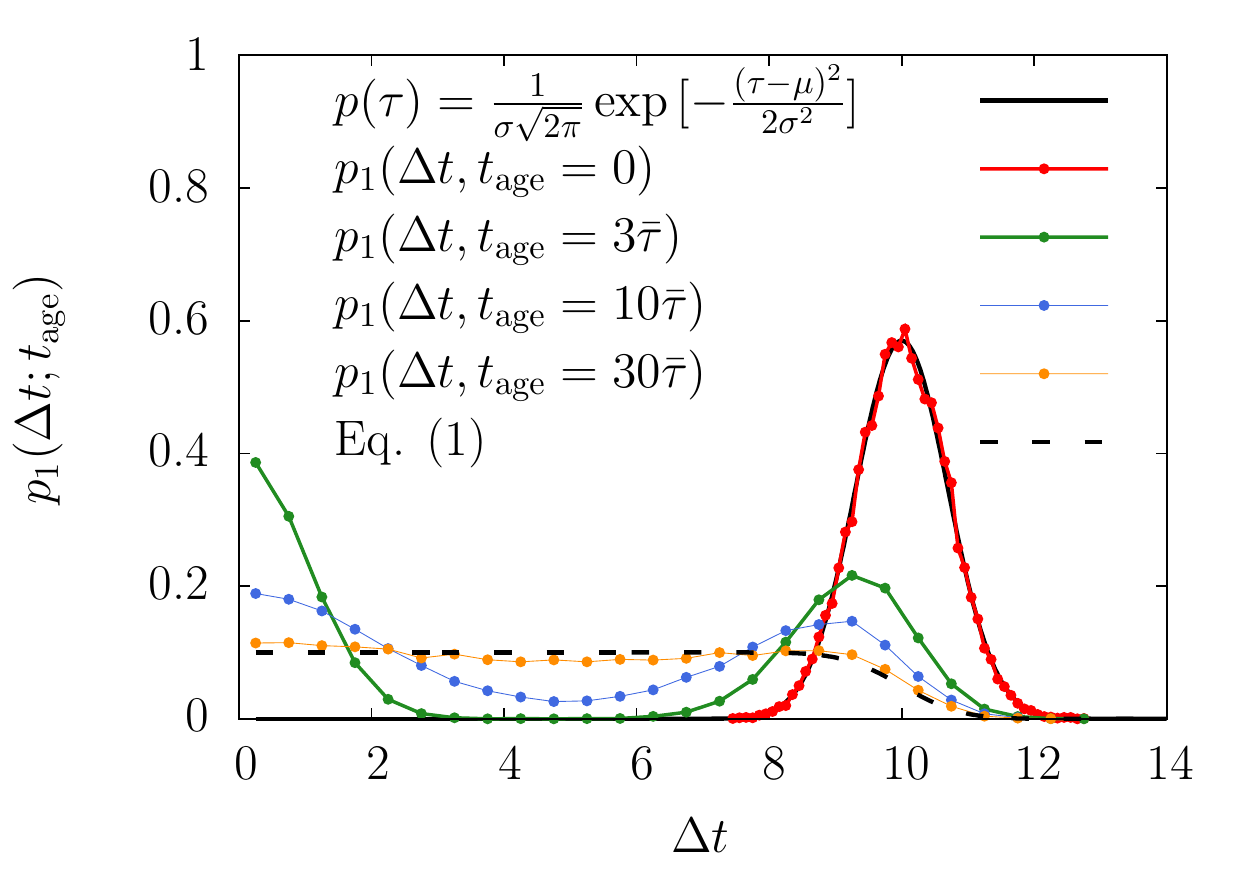}
(b)\includegraphics[width=0.4\textwidth]{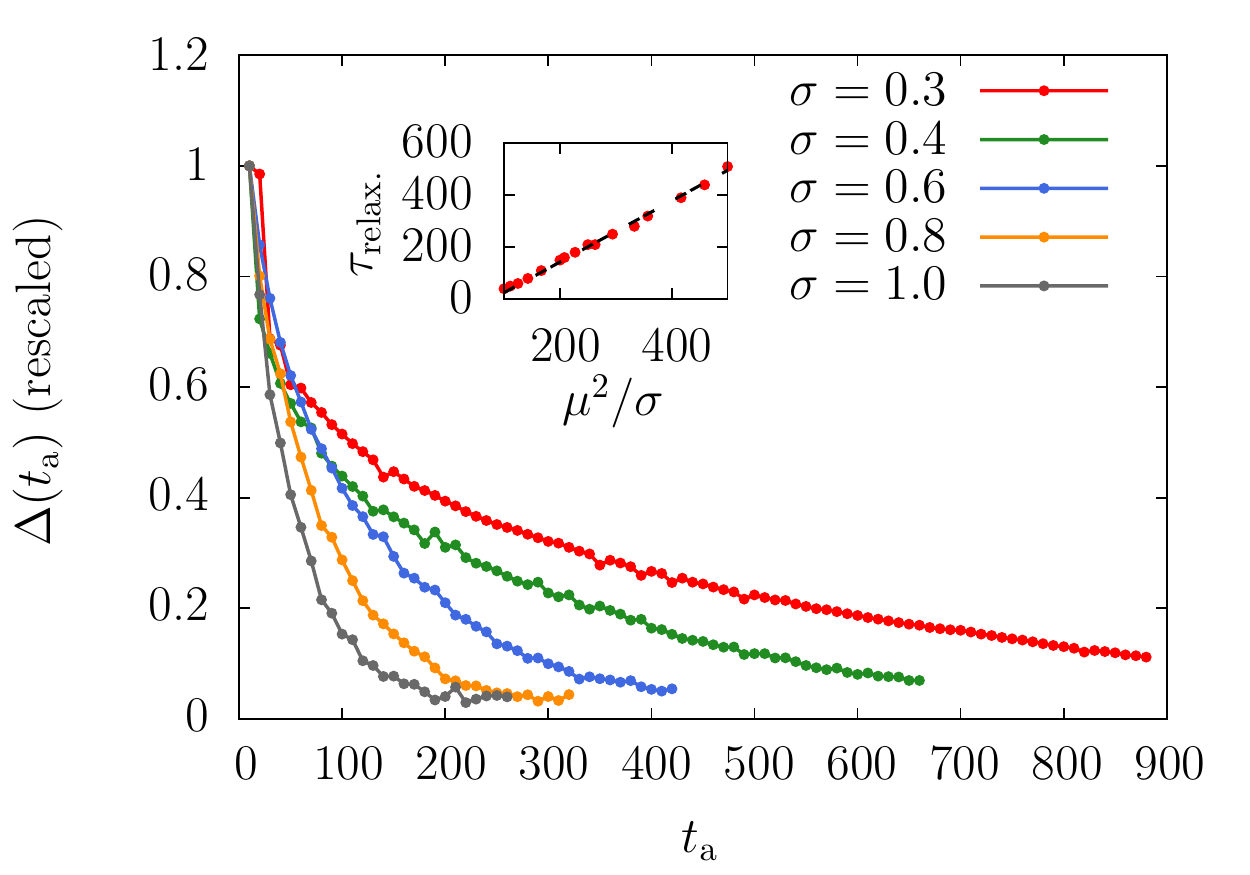}
\caption[]{(a) First passage time, $p_1(\Delta t)$, for various $\ta$, obtained from a CTRW via the procedure described in \figref{fig:CTRW-schematic}. The corresponding persistence time distribution is drawn from an age-independent Gaussian function with the mean value $\overline\tau=10$ and standard deviation $\sigma=0.7$. Due to the synchronization at $t=0$, the obtained first passage time distribution for $\ta=0$, is identical to $p(\tau)$. In the limit of large $\ta$, on the other hand, the memory on synchronization is lost and $p_1(t)$ is well described by \equref{eq:p1-equil}. This limiting case is reproduced by randomizing the time origin, i.e., by considering a shift $t_0$, randomly chosen in $[0,\; \tau] $, to the value which is drawn from $p(\tau)$ for the first jump. By doing so, we cancel the memory effect at once without the need for long CTRW trajectories.  (b) The zeroth moment of the difference, 
$\Delta (\ta):=\int_{0}^{\infty} ||p_1(s; \ta)-\lim_{\ta\rightarrow\infty} p_1(s; \ta)|| ds$ versus $\ta$. As expected, narrower $p(\tau)$ lose their memory of synchronization more slowly. 
The relaxation time, $\tau_{\text{relax}}$, is determined by the condition that $\Delta (\ta=\tau_{\text{relax}})=\Delta(0)/e$. The inset shows the thus obtained relaxation time. The dashed line is the prediction of  \equref{tdesynch}.}
\label{fig:p1-evolution}
\end{figure}

Indeed, in a CTRW, one successively draws waiting times $\tau_i$, $i$ being an integer index. Then, given the aging time $\ta$, the first passage time is determined as the time between $t=\ta$ and the first event occurring at $t_i>\ta$ (\figref{fig:CTRW-schematic}). One obtains $p_1(\Delta t)$ by repeating this procedure. Obviously, for $\ta=0$, the thus obtained distribution function, $p_1$, will be identical to the persistence time distribution, $p$. However, as $\ta$ grows, $p_1$ will progressively deviate from $p$ until it becomes fully independent of $p$ in the limit of large aging times. In this limit, one expects the thus obtained $p_1$ to obey \equref{eq:p1-equil}. This behavior is evidenced in \figref{fig:p1-evolution}a.

The characteristic  time needed for the decay of the correlations between $p_1$ and $p$ depends on the function $p(\tau)$, e.g. its mean and standard deviation. For example, for a narrower distribution, a larger number of jumps is needed to randomize the dynamics; in the limiting case of the Dirac delta function, by synchronizing particles at $t=0$, they remain synchronized for all times and perform their jump simultaneously. For two random walkers synchronized at time zero, the number of jumps needed to lose the memory of synchronization (i.e., for one of them to get ahead of the other by one jump) is of the order of $\overline\tau/\sigma$, where $\overline\tau$ is the average time between two successive jumps, and $\sigma=\sqrt{\overline{\tau^{2}}-\overline{\tau}^2}$ is the mean standard deviation of the time distribution (i.e., a measure of the fluctuations of $\tau$). The time required for randomization is estimated by multiplying the number of jumps $\overline\tau/\sigma$ necessary for desynchronization with the average waiting time $\overline\tau$, 
\begin{align} \label{tdesynch}
\tau_\mathrm{desync.} \sim \frac{\overline\tau^2}{\sigma}.
\end{align}
The validity of this estimate is assessed  for the CTRW model with the Gaussian PDF, as shown in \figref{fig:p1-evolution}b.


\section{Conclusion}\label{Conclusion}
This letter addresses the close connection between the first passage time, $p_1$, and the persistence time, $p$, distribution functions in glass-forming systems subject to physical aging. It is shown that, in contrast to a number of recent works, both the first passage time and the persistence time distribution functions show aging behavior. It is argued here that resolving aging effect in $p$ requires longer simulations due to the longer aging time associated with the detection of two successive events, as compared to $p_1$ which requires the detection of the first of these two events only. As a result, when determined with the same numerical effort, $p(\tau)$ shows weaker age-dependence than $p_1$.  This age-dependence of the persistence time distribution is sensitive to the details of the algorithm used to extract it from particle trajectories. By updating the reference point in event detection algorithm and accounting for the event specific aging time, we uncover the age dependence of $p(\tau)$, hidden to previous studies. The apparent aging effects in continuous time random walk models which make use of an age-independent waiting time distribution is also investigated. It is shown that the main reason for age-dependence of $p_1$ within CTRW lies in the implicit synchronization of trajectories within this model, and is thus genuinely different from the observed aging in the MD simulations, the latter reflecting the evolution of the system towards an equilibrium state. Thus, when studying aging phenomena within a CTRW model, one must remedy for the implicit synchronization of the relaxation events---e.g., via randomizing the time origin---so that one can focus on physically relevant aging processes.

\acknowledgments
\emph{Acknowledgments.---} We thank Jean-Louis Barrat, Andreas Heuer and J\"org Rottler for useful discussions. N.H.S. acknowledges financial support by the AICES grant No GSC 111. F.V. on behalf of ICAMS acknowledges funding from its industrial sponsors, the state of North-Rhine Westphalia, and the European Commission in the framework of the European Regional Development Fund (ERDF).

\bibliographystyle{prsty}
\bibliography{EPL_G34867_revised_26_08_2015}

\end{document}